\documentclass[preprint]{aastex}
\usepackage{amsbsy} 

\newcommand{\Msun}{M_\odot}

\slugcomment{submitted to MNRAS}

\shorttitle{Molecular Line Profiles from Contracting Dense Cores}
\shortauthors{Stahler}

\begin{document}

\title{Molecular Line Profiles from Contracting Dense Cores}

\author{Steven W. Stahler\altaffilmark{1} and Jeffrey J. Yen\altaffilmark{2}}

\altaffiltext{1}{Astronomy Department, University of California,
Berkeley, CA 94720}
\altaffiltext{2}{Physics Department, Stanford University,
Stanford, CA 94305}

\email{Sstahler@astro.berkeley.edu}

\begin{abstract}
We recently proposed that molecular cloud dense cores undergo a prolonged
period of quasi-static contraction prior to true collapse. This theory could
explain the observation that many starless cores exhibit, through their 
spectral line profiles, signs of inward motion. We now use our model, 
together with a publicly available radiative transfer code, to determine the 
emission from three commonly used species - N$_2$H$^+$, CS, and HCN. A 
representative dense core of $3\,\,\Msun$ that has been contracting for 1~Myr 
has line profiles that qualitatively match the observed ones. In particular, 
optically thick lines have about the right degree of blue-red asymmetry, the
empirical hallmark of contraction. The \hbox{$J\,=\,2\rightarrow\,1$} 
rotational transition of CS only attains the correct type of profile if the 
species is centrally depleted, as has been suggested by previous studies. These
results support the idea that a slow, but accelerating, contraction leads to 
protostellar collapse. In the future, the kind of analysis presented here can 
be used to assign ages to individual starless cores.
\end{abstract}


\section{Introduction}

Over the last decade, astronomers have carefully scrutinized the large
fraction of molecular cloud dense cores that contain no point sources of 
radiation, in order to understand more fully the onset of star formation. It 
was already established that the gross properties of these objects differ 
little from their counterparts with internal stars \citep{lm99}, except perhaps
for a mass that is slightly higher on average \citep[][Fig.~3]{jm99}. More 
detailed spectroscopic studies of molecular line emission revealed that many
starless cores exhibit signs of inward contraction \citep{wi99,lmt99,ge00,
lmt01,sc07}. The hallmark of this so-called infall signature is an asymmetric 
emission line, often with a self-absorption dip, that is skewed to the blue. 

There is a variety of line profiles seen in dense cores, sometimes even in
the same object, and the profile just described is by no means universal. It is
preferentially found in starless cores with a higher central column density and
optical depth \citep[e.g.][]{lmt99,ge00}. Such cores also exhibit chemical 
abundances that differentiate them from lower-density objects 
\citep{c05,kc08}. The kinematic signature is especially common in more evolved 
dense cores that contain Class~0 and Class~I sources, i.e., very young, 
embedded stars \citep{me00}. These facts, together with the simple and 
compelling infall interpretation of the asymmetric profiles \citep{m96}, have 
convinced most researchers that we are witnessing a type of contraction that 
precedes the free-fall collapse onto a protostar. The contraction hypothesis is
especially convincing for starless cores, where there is no possibility of 
confusion from stellar outflows.

What is the physical cause of this motion? The observed magnitude of the 
infall velocity is typically a few tenths of the sound speed. This finding 
rules out ambipolar diffusion as the underlying process, since the 
characteristic ion-neutral drift speed is much smaller 
\citep[e.g.][]{cb01}. Conversely, the subsonic level of the velocity, together
with the statistical prevalence of starless cores \citep[e.g.][]{f09}, 
indicate that such objects are relatively long-lived \citep{ki05}, and thus are
unlikely to be in a true state of collapse, as proposed by \citet{kf05}. 

In a recent contribution \citep[][hereafter Paper~I]{sy09} we introduced a new 
model for the contraction. Starless cores, like all self-gravitating objects, 
undergo bulk oscillations. Indeed, there are now several cases where 
spectroscopic observations indicate that such oscillations are occurring 
\citep{r06,b07,v08}. For an object that has evolved to the point of collapse, 
by whatever means, the frequency of the lowest normal mode vanishes. We 
studied, using perturbation theory, how a cloud subject to such a frozen (more 
commonly called `neutral') mode slowly evolves toward full-blown collapse. We 
found that this previously unrecognized phase of quasi-static contraction lasts
a considerable period (of order $10^6$~yr), during which time the cloud attains
internal speeds similar to those observed (see Fig.~4 of Paper~I). 

Encouraged by this result, we now seek to forge a stronger link to the 
observations. We calculate, using a publicly available radiative transfer code,
the emitted line spectrum from our model cloud. We choose three molecules
that are among the most commonly employed in the spectroscopic studies: 
N$_2$H$^+$, CS, and HCN. Following the early surveys of \citet{tt77} and
\citet{w92}, N$_2$H$^+$ is often used as a tracer of dense, relatively
quiescent cloud gas. Its rotational lines are optically thin in low-mass 
dense cores. Thus, they do not exhibit either self-absorption or blue-red 
asymmetry, but serve principally as a gauge of excitation temperature and as a
comparison benchmark with lines that are sensitive to internal cloud motion.
The rotational lines of CS are often optically thick, and can have asymmetric 
profiles that have long been interpreted kinematically \citep{w86}. On the 
other hand, the infall signature is relatively weak in starless cores, in part 
because the polar molecule CS is depleted at high density from the gas phase 
onto grain mantles \citep{bl97}. Finally, the \hbox{$J\,=\,1\rightarrow\,0$} 
rotational transition of HCN has the strongest infall signature of all, and the
molecule appears to suffer little depletion \citep{a05,s07}.

Although our model cloud is a highly idealized isothermal sphere, the results
of this study further lend credence to the underlying dynamical picture. For a
representative 3\,\,$\Msun$ cloud that has been contracting for 1~Myr, we find
that all the calculated line profiles resemble those observed, at least in an
average sense. Interestingly, the CS profile is only acceptable once we include
central depletion of the molecule. In Section~2 below, we first describe our
implementation of the radiative transfer program. Section~3 then summarizes 
observations to date of the three molecules in starless dense cores. We 
describe, in Section~4, how the emergent line profiles alter when we change our
physical input parameters. Judging from these trends, we then select a 
canonical model cloud and compare its emission spectrum to the available data. 
Finally, Section~5 discusses limitations of the dynamical theory and the 
prospects for improvement. We also indicate how the model, even in its current,
simplified form, can be used to assign contraction ages to individual
starless cores.   

\section{Calculating Line Emission}

For each selected cloud model, we solve the equation of transfer numerically, 
using the code RATRAN \citep{h00}. The basic strategy underlying this 
code is to choose randomly a set of cells outside the cloud that generate 
photons. These photons propagate through the cloud interior, while the 
molecular level populations are temporarily held fixed. The statistical 
equilibrium of the levels is then solved separately, whereupon photon 
propagation is reinitiated. The accelerated lambda iteration scheme introduced
by \citet{rh91} ensures rapid convergence of the radiation field and level
populations. Treatment of non-isotropic scattering is cumbersome in this 
approach, so RATRAN does not include this effect. Such a simplification is 
acceptable in the infrared and millimeter regime of interest. 

The code must be supplemented by tables of molecular data, containing energy 
levels, degeneracies, Einstein A-values, and collisional rate coefficients. 
Here, we utilized the Leiden Atomic and Molecular Database (LAMDA). 
\citet{s05} describe the construction of the database and its use. These tables
include the full multiplets of rotational lines created by hyperfine splitting.
In fact, recent observational studies of dense core kinematics have focused on 
individual hyperfine lines, as we will describe.

In order to display the line spectra, we utilized the MIRIAD software package
\citep{s95}. We placed our cloud at a representative distance of 140~pc, and 
``observed'' it with a Gaussian beam of angular size $60^{\prime\prime}$. For
comparison, our canonical cloud has a physical diameter of 0.27~pc, or
392$^{\prime\prime}$ at the selected distance. In constructing spectra, we
assumed a velocity resolution of 0.01~km~s$^{-1}$.

We tested the radiative transfer calculation by first reproducing prior results
for line profiles in a collapsing cloud. An early and influential work was that
of \citet{z92}. This study compared the self-similar collapse models of 
\citet{s77} with that derived independently by \citet{L69} and \citet{p69}. 
\citet{z92} calculated, for both models, CS rotational line profiles at various
times, both before and after formation of the central protostar.\footnote{The
original Larson-Penston model only considered collapse up to the point of
protostar formation. \citet{z92} utilized the calculation of \citet{h77} to
extend the flow past this point. Conversely, \citet{s77} only considered times
{\it after} protostar formation. For the earlier epoch, \citet{z92} employed a
sequence of hydrostatic spheres of increasing central density.}   

In testing both self-similar collapse models, \citet{z92} assumed a gas kinetic
temperature of 10~K. Following \citet{ls89}, he augmented the Shu model with
a turbulent velocity of the form
\begin{equation}
 v_{\rm turb} \,=\, \sqrt{\kappa/\rho} \,\,,
\end{equation}
where $\kappa$ is a constant and $\rho$ is the local cloud density. The
constant was determined empirically by \citet{mg88} to have the value
\hbox{$\kappa\,=\,1.1\times 10^{-11}\,\,{\rm g}\,\,{\rm cm}^{-1}\,\,
{\rm s}^{-2}$}. \citet{z92} argued that this figure was likely to be an
overestimate, since the observed ``turbulence'' could have included infall 
motion for dense cores containing stars. Accordingly, he used a value of 
$\kappa$ smaller by a factor of 0.25. Finally, he set the CS abundance, i.e.,
the number density of the molecule relative to hydrogen, to $4\times 10^{-9}$, 
following \citet{f89}.

We provisionally adopted the same assumptions as \citet{z92} and recalculated,
using RATRAN, CS line profiles from the Shu model for the period after 
protostar formation. All lines were viewed through the cloud center. The 
RATRAN code allows one to select $N_{\rm tot}$, the total number of shells in 
the cloud, as well as $r_{\rm out}/r_{\rm in}$, the ratio of outer to inner 
shell radii. We adjusted both numbers until we obtained the best match with 
the \hbox{$J\,=\,1\rightarrow 0$}, \hbox{$J\,=\,2\rightarrow 1$}, and 
\hbox{$J\,=\,3\rightarrow 2$} profiles  shown in Figure~3d of \citet{z92}. By 
choosing \hbox{$N\,=\,100$} and \hbox{$r_{\rm out}/r_{\rm in}\,=\,500$}, we 
obtained an excellent fit to all profiles for the four times considered by 
\citet{z92}. Agreement did not improve when either $N_{\rm tot}$ or 
$r_{\rm out}/r_{\rm in}$ was increased further. For example, the peak antenna
temperatures $T_A$ changed by less than 0.1~\%. We therefore retained these 
values of $N_{\rm tot}$ and $r_{\rm out}/r_{\rm in}$ in our subsequent 
calculations. 

Turning to starless dense cores, we continued to use a gas kinetic
temperature of 10~K, corresponding to a thermal speed of
\hbox{$a_T\,=\,0.2\,\,{\rm km}\,\,{\rm s}^{-1}$}. This temperature is on the
low end of the range found empirically by \citet[][Figure 2]{jm99}. In any 
case, it is important to note that purely thermal line profiles are never seen,
even at the centers of the most quiescent objects \citep{bg98}, and the 
prescription for turbulence in equation~(1) continues to provide a 
useful representation of the observations. In light of the comments by 
\citet{z92}, we view $\kappa$ as a free parameter, whose upper limit is likely 
to be the value advocated by \citet{mg88}. It is straighforward to incorporate 
turbulence into RATRAN, which allows a variable degree of intrinsic line 
broadening in each mass shell of the cloud.

An additional free parameter is inherent in the theoretical model of Paper~I.
We envisioned the cloud undergoing a global oscillation, characterized by a
nondimensional amplitude $\epsilon$. For our standard initial conditions, 
$\epsilon$ gives the fractional deviation of the density from its unperturbed,
equilibrium distribution (see eq.~(17) and the discussion in \S 4.2). The
numerical integration of Paper~I used a relatively large amplitude of 
\hbox{$\epsilon\,=\,0.2$}, motivated by observation and analysis of the 
starless core B68 \citep{k06}. Here, we reset $\epsilon$ to the more modest,
and perhaps more typical, value of 0.1. We comment later on how changing
$\epsilon$ affects the emergent line profiles.

With these assumptions, the only remaining free parameters are the cloud mass
$M$ and the time of observation $t$. The zero point for $t$ is the onset of
quasi-static contraction. Both quantities must be related to the corresponding
nondimensional ones, $\tilde M$ and $\tilde t$, used in Paper~I. Our model
cloud has a fixed nondimensional mass, \hbox{${\tilde M}\,=\,4.19$}, where
the number results from numerical integration of the isothermal Lane-Emden
equation. Combining equation~(9) and (12) of Paper~I, we find that $\tilde M$
and $\tilde t$ are related by
\begin{equation}
 {\tilde t} \,=\, {{{\tilde M}\,a_T^3\,t}\over{G\,M}} \,\,.
\end{equation}
Since $\tilde M$ and $a_T$ are either known or preset, the equation tells us, 
for any $M$ and $t$, the appropriate $\tilde t$ to use from the model sequence.
We further read off, at this point in the sequence, the distribution of the 
nondimensional density and velocity as functions of radius. The equations of 
\S 2.2 in Paper~I tell us how to translate the quantities into dimensional 
form.\footnote{These equations contain the external pressure $P_e$. 
Equation~(9) of Paper~I allows us to solve for $P_e$, given the dimensional 
cloud mass $M$.}

\section{Observations of the Molecules} 

\subsection{N$_2$H$^+$}

After surveys had established the presence of N$_2$H$^+$ in star-forming
regions, targeted observations were pursued in dense cores, including those 
already detected in NH$_3$ \citep{c95,b98,c02}. It was found that the molecule
selectively traces dense and quiescent gas, and is indeed absent in hotter and
shocked material. The \hbox{$J\,=\,1\rightarrow 0$} rotational transition, 
which has 7 hyperfine components, is usually optically thin in starless cores,
but not always so. For example, the well-studied core L1544 shows an asymmetric
profile toward the cloud center, extending over a region some 1400~AU in radius
\citep{t98,wi99}.

\citet{lmt99} conducted the first largescale, systematic observations of 
N$_2$H$^+$ in starless dense cores. The 220 sources in their catalog had been 
selected primarily on the basis of optical obscuration. For the 72 objects 
detected in N$_2$H$^+$, detailed study was made of the relatively isolated
\hbox{${F_1 ,F}\,=\,{0,1} \rightarrow {1,2}$} hyperfine line at 93.176~GHz.
This line almost always has a Gaussian profile, with no self-absorption or 
asymmetry. The mean measured peak intensity was 
\hbox{$T_A\,=\,0.19\,\pm\,0.11\,\,{\rm K}$}, and the FWHM was
\hbox{$\Delta V_{{\rm N}_2 {\rm H}^+} \,=\, 0.35\,\pm\,0.14\,\,{\rm km}\,\,
{\rm s}^{-1}$}. \citet{lmt01} followed up these single-pointing observations 
with spatial maps of many individual cores. The N$_2$H$^+$ emission is compact,
with a well-defined central peak. Diameters of the regions within the 
half-maximum contours range from 0.05 to 0.13~pc.  

\subsection{CS}

These properties of N$_2$H$^+$ emission are in sharp contrast to those of the
\hbox{$J\,=2\,\rightarrow\,1$} rotational line of CS at 97.981~GHz. 
\citet{lmt99} found, in the 163 starless cores for which the line was detected,
that a large fraction of the profiles were asymmetric and skewed toward the
blue, either with two distinct peaks or, more commonly, a blue peak and a red
shoulder (see their Figure~1). Detailed profiles vary from source to source, 
and there are even a few examples of double-peaked profiles stronger in the 
red. The subsequent mapping survey of \citet{lmt01} showed that the region of 
CS asymmetry is relatively diffuse, and the emission is frequently depressed 
just where the  N$_2$H$^+$ intensity peaks \citep[see also][]{t02}.

With both the optically thin (N$_2$H$^+$) and optically thick (CS) lines in 
hand, one may quantify the degree of blue-red asymmetry of the latter. 
\citet{m97} introduced $\delta V_{\rm CS}$, the normalized velocity shift of
CS. This quantity is defined as
\begin{equation}
 \delta V_{\rm CS} \,\equiv\, {{V_{\rm CS}\,-\,V_{{\rm N}_2 {\rm H}^+}}\over 
{\Delta V_{{\rm N}_2 {\rm H}^+}}} \,\,.
\end{equation}
Here, $V_{\rm CS}$ and $V_{{\rm N}_2 {\rm H}^+}$ are the peak, line-of-sight
velocities of the two species. The peak velocity of N$_2$H$^+$ is assumed to 
represent the systematic motion of the cloud as a whole. In the single-pointing
survey of \citet{lmt99}, $\delta V_{\rm CS}$ was usually negative, signifying
that the profile peak is blueward of line center. The mean value of 
$\delta V_{\rm CS}$ in this sample was $-0.24\,\pm\,0.04$.

\subsection{HCN}

The \hbox{$J\,=\,1\,\rightarrow\,0$} rotational transition of HCN is the latest
diagnostic tool for studying the kinematics of starless cores. As mentioned
previously, the molecule suffers little depletion, at least for central
densities less than \hbox{$10^5\,\,{\rm cm}^{-3}$} \citep{a05}. Integrated
intensity maps show that the HCN emission region is roughly coextensive with
that of N$_2$H$^+$ \citep{s04}. The \hbox{$J\,=\,1\,\rightarrow\,0$} transition
consists of three hyperfine lines of differing optical depth:
$F\,(0\!-\!1)$ at 88.634~GHz, $F\,(2\!-\!1)$ at 88.632~GHz, and 
$F\,(1\!-\!1)$ at 88.630~GHz, Under optically thin, LTE conditions, these would
have intensity ratios of $1:5:3$. Of course, it is the fact that these lines 
are optically thick in starless cores that makes them useful kinematic probes. 

\citet{s07} conducted single-pointing observations for HCN in 85 starless 
cores, netting 64 detections. The profiles of all three hyperfine lines are
distinctly non-Gaussian, displaying either two peaks with a self-aborption
dip or a single peak with a shoulder. Not surprisingly, the central dip is
strongest in the optically thickest $F\,(2\!-\!1)$ line, which also shows the
greatest blue-red asymmetry. But a central result of this study is the
heterogeneity of the profiles. While 43~percent of the sources have at least
one or two double-peaked hyperfine lines skewed to the blue, 33~percent have
blueward and redward asymmetries mixed in the three lines. 

Nevertheless, the overall tendency is for profiles indicative of contraction, 
rather than expansion. In analogy with equation~(3), \citet{s07} defined 
$\delta V_{\rm HCN}$, a normalized velocity shift with respect to N$_2$H$^+$.
For each hyperfine line, the large majority of sources are observed to have
negative $\delta V_{\rm HCN}$ (see their Figure~4). The actual magnitudes of
$\delta V_{\rm HCN}$ are larger than for CS. The mean values are
$-0.51\,\pm\,0.06$, $-0.41\,\pm\,0.10$, and $-0.59\,\pm\,0.08$ for
$F\,(0\!-\!1)$, $F\,(2\!-\!1)$, and $F\,(1\!-\!1)$, respectively. Moreover,
the magnitudes are tightly correlated, in the sense that a source with 
greater $\vert\delta V_{\rm HCN}\vert$ in the $F\,(0\!-\!1)$ line also has a
greater shift in $F\,(2\!-\!1)$ and $F\,(1\!-\!1)$. Within each source,
however, there is no clear pattern for the relative magnitudes of the
velocity shifts for the three lines, nor for the relative intensities of the 
lines. In other words, starless cores frequently exhibit HCN ``intensity 
anomalies'' with respect to LTE, a phenomenon that has long been noted and 
studied in other molecular clouds \citep[e.g.,][]{g93}.

Because of the range in optical depths in the three lines, the profiles of
HCN are potentially useful for tracing the full velocity structure of a 
dense core. \citet{l07} constructed single-dish spatial maps of emission for 
two starless cores, L694-2 and L1197. To complement their observations, they
also performed a radiative transfer calculation in spherical cloud models
with adjustable density and velocity profiles. Interestingly, monotonic 
velocity profiles do not match the data. For both starless cores, \citet{l07}
found that the inferred contraction speed begins relatively small at the cloud
edge, reaches a maximum about halfway inward in radius, and thereafter declines
toward the cloud center. This type of velocity profile emerges naturally from 
our dynamical model (Paper~I; Figure~4).

\section{Numerical Results}

\subsection{Trends in Calculated Line Profiles}

After setting the oscillation amplitude to \hbox{$\epsilon\,=\,0.1$}, we next
gauge how the line profiles, still viewed through the cloud center, change as 
the other parameters are varied. These trends enable us to seek a model whose 
lines best match the data in an average sense. Recall that our remaining free 
parameters are: $\kappa$, the turbulence coefficient in equation~(1), the cloud
mass $M$, and the time $t$ since the start of inward motion. The updated 
molecular abundances we used in the transfer code were: 
\hbox{$X\,({\rm N}_2{\rm H}^+)\,=\,7\times 10^{-10}$} \citep{b98}; 
\hbox{$X\,({\rm CS})\,=\,4\times 10^{-9}$} \citep{t02}; and
\hbox{$X\,({\rm HCN})\,=\,3\times 10^{-9}$} \citep{f91,l04}.

We first explored the effect of altering the cloud mass at fixed $\kappa$ and
$t$. Following \citet{lmt99}, we focused on the 
\hbox{${F_1 ,F}\,=\,{0,1} \rightarrow {1,2}$} hyperfine line of N$_2$H$^+$.
\footnote{The N$_2$H$^+$ table of LAMDA lists 3 lines at 93.176~GHz 
corresponding to the transitions
\hbox{${F_1 ,F}\,=\,{0,1} \rightarrow {1,2}$},
\hbox{${F_1 ,F}\,=\,{0,1} \rightarrow {1,1}$}, and
\hbox{${F_1 ,F}\,=\,{0,1} \rightarrow {1,0}$}. The lower levels are also listed
as being degenerate in Table~1 of \citet{d05}. Assuming the three individual  
lines cannot be resolved in practice, we added their intensities.} As $M$ is 
raised, the N$_2$H$^+$ line remains Gaussian, and is therefore still optically 
thin. Of course, the total emission increases with $M$, so that 
$\Delta V_{{\rm N}_2{\rm H}^+}$ also rises. 

In CS and HCN lines exhibiting self-absorption, these dips eventually disappear
with increasing cloud mass. This tendency is easy enough to understand. 
According to equation~(2), an increase of $M$ at fixed $t$ necessitates a lower
$\tilde t$. We are viewing our fiducial, contracting cloud at an earlier
nondimensional time, when the cloud radius $\tilde r$ was larger. But 
equations~(9), (10), and (12) of Paper~I imply that  
\begin{equation}
M\,R^{-2} \,=\, {{{\tilde m}\,{\tilde t}}\over{{\tilde r}^2}}
\, {a_T\over{G\,t}} \,\,.
\end{equation}
Since ${\tilde t}/{\tilde r}^2$ falls, the column density through the cloud
{\it decreases} with increasing $M$. Many observed lines, especially those of 
HCN, show clear self-absorption. Thus, the sensitivity of the calculated
profiles effectively sets a ceiling on the allowable range of cloud masses.

A similar disappearance of the self-absorption dips in CS and HCN occurs when
we fix $M$ and $t$, but increase $\kappa$. A larger value of $\kappa$ 
corresponds to a greater velocity dispersion throughout the cloud. The gas
consequently has a higher, overall excitation temperature, so there is less of
the absorption that creates central line dips. By the same token, the
optically thin N$_2$H$^+$ line broadens, i.e., 
$\Delta V _{{\rm N}_2{\rm H}^+}$ increases. Both trends help us to set bounds 
on $\kappa$.

Finally, we may fix $M$ and $\kappa$, and consider various evolutionary times
$t$. At greater $t$, the maximum infall speed within the cloud increases. Thus,
the blue/red asymmetry of all optically thick lines becomes more pronounced.
Quantitatively, both ${\delta V}_{\rm CS}$ and ${\delta V}_{\rm HCN}$ which 
are both negative, become larger in magnitude. Figure~1 shows that the overall
emission in N$_2$H$^+$ increases with time, a reflection of the higher cloud
density. The line peak grows faster than its width, with the result that
$\delta V _{{\rm N}_2{\rm H}^+}$ declines. In the figure, the FWHM is 
\hbox{$0.24~{\rm km}\,\,{\rm s}^{-1}$} at the initial instant, and
\hbox{$0.21~{\rm km}\,\,{\rm s}^{-1}$} at the last time shown 
(\hbox{$t\,=\,1\,\,{\rm Myr}$}).

\subsection{Canonical Model}

Rather than attempt to find, within the model parameter space, matches to
individual dense cores, we sought first a combination of $M$, $t$, and 
$\kappa$ that reproduces the general, observed characteristics of the various 
line profiles, as summarized in \S 3. Some experimentation was required, in 
light of the constraints just given. However, a suitable parameter set quickly 
emerged. If we set \hbox{$M\,=\,3\,\,\Msun$}, \hbox{$t\,=\,1\,\,{\rm Myr}$}, 
and 
\hbox{$\kappa\,=\,5.5\times 10^{-12}\,\,{\rm g}\,\,{\rm cm}\,\,{\rm s}^{-2}$}
(i.e., 0.5 times the original value found by \citet{mg88}), then the profiles
are generally consistent with the data.

Following our previous discussion, it is the magnitude of the self-absorption 
dip that is most sensitive to variations in these parameters. We may 
accordingly gauge the acceptable range of parameter values by noting when the
dips depart substantially from those in the canonical model. Holding $t$ and
$\kappa$ fixed while allowing $M$ to range from 2.7 to $3.2\,\,\Msun$, the
magnitude of the dip in the $F\,(2-1)$ line of HCN changed by less than
25~\% in either direction. By the analogous measure, $t$ can vary from 0.8 to
1.2~Myr, and $\kappa$ from 3.3 to 
\hbox{$8.8\times 10^{-12}\,{\rm g}\,{\rm cm}\,{\rm s}^{-1}$}. 

The topmost curve in Figure~1 shows the calculated N$_2$H$^+$ profile for the 
canonical model. As in nearly all observed cases, the line is symmetric and
Guassian. Our FWHM of 0.21~km~s$^{-1}$ is on the low end of the observed 
range, as would be expected given our assumed kinetic temperature of 10~K. 
Conversely, our T$_{\rm A}$-value of 0.33~K is higher than the average. Here, 
of course, the beam efficiency (assumed to be unity in RATRAN) comes into play,
as do the various distances to the observed dense cores.

Figure~2 (dashed curve) shows the emergent profile of CS through the center of 
the canonical cloud model. Because of higher optical depth, the line is 
markedly asymmetric, as the data also show. However, while the sense of the 
asymmetry in this calculated profile is correct, the deep, self-absorption dip 
is rarely seen (compare Figure~1 of \citet{lmt99}). What is observed instead 
is a relatively broad, redshifted shoulder joining onto a blueshifted peak. 

We may recover this shape by introducing depletion of the molecule, presumably 
caused by adsorption onto grain surfaces. \citet{t02} invoked such depletion to
improve the observational match of their CS spatial intensity maps. Following 
these authors, we assume a centrally depressed, radial profile for the relative
number density:
\begin{equation}
X\,(r)\,=\, X_\circ\,\,{\rm exp}[-n(r)/n_d]  \,\,.
\end{equation}     
Here, $X_\circ$ is the usual interstellar value of $X\,({\rm CS})$, $n(r)$ is
the cloud number density, and the threshold density $n_d$ is a free parameter.
For the solid profile in Figure~2, we used 
\hbox{$n_d\,=\,1\times 10^4\,\,{\rm cm}^{-3}$}. As a comparison, Table~4 of
\citet{t02} lists $n_d$-values ranging from 1 to 
$5\times 10^4\,\,{\rm cm}^{-3}$ for five starless cores. Our calculated 
relative velocity shift, \hbox{$\delta V_{\rm CS}\,=\, -0.24$}, is equal to 
the mean observed \citep{lmt99}.  

The three hyperfine lines of HCN (Figure~3) exhibit more diverse profile
shapes. The optically thinnest $F\,(0-1)$ transition has a blue peak/red 
shoulder profile, similar to that in CS. The other two lines exhibit clear 
self-absorption, with the deepest trough being found in the optically
thickest, $F\,(2-1)$ line. Scanning Figure~1 of \citet{s07}, we see
that this pattern is commonly seen (e.g., L63, L1197), but is by no means
universal (e.g., L1544). Our calculated relative velocity shifts, $\delta V$,
for the (0-1), (2-1), and (1-1) lines are -0.33, -0.81, and -0.67, 
respectively. As mentioned earlier, observations give similar values for all
three lines, with a mean of about -0.5. We are further encouraged by this
qualitiative agreement.

We stress that calculated line profiles are quite sensitive to the interior 
dynamics of the cloud, and thus do provide a worthwhile test of the underlying 
theory. To further illustrate this point, we recalculated all profiles using 
the collapse model of \citet{s77}. We again set 
\hbox{$t\,=\,1\times 10^6\,\,{\rm yr}$}, where the zero of time now refers
to the start of inside-out collapse. For \hbox{$T\,=\,10\,\,{\rm K}$}, the
mass of the central protostar in the model is $2\,\,\Msun$. In calculating
profiles, we used the same velocity dispersion parameter,
\hbox{$\kappa\,=\,5.5\times 10^{-12}\,\,{\rm g}\,\,{\rm cm}\,\,{\rm s}^{-2}$},
as in our canonical model. 

All calculated line profiles are now far broader than those observed. For 
example, the N$_2$H$^+$ profile, while remaining symmetric, is distinctly
non-Gaussian, with a shallow, central depression and wings extending to 
\hbox{$\pm1\,{\rm km}\,\,{\rm s}^{-1}$} on either side of line center. 
Optically thick lines have profiles that are much too asymmetric. Our point 
here is {\it not} that the Shu model is a poor representation of starless 
cores, which was to be expected. The message is rather that the gas motions 
induced by a central protostar of solar-type mass are qualitatively different 
from those in our slowly contracting, starless cloud, and in a manner clearly 
reflected in the molecular line profiles.  

\subsection{Determining $\epsilon$ and $t$}

Our canonical value of the oscillation amplitude $\epsilon$, while guided by 
the scant observational data available, is still rather arbritrary. How, in the
future, can we pin down this parameter when we examine specific dense
cores? Even more importantly, will we be able to gauge the cloud's contraction
age through comparison between real and model-generated line profiles?

These questions are related. We remarked in Paper~I that an increase of 
$\epsilon$, if accompanied by a {\it decrease} in the evolutionary time, could 
leave the cloud largely unchanged. If that were the case, it would be difficult
to ascertain $\epsilon$ and $t$ individually. Specifically, our cloud model, at
a given $\epsilon$ and nondimensional time $\tilde t$, attains nearly the same 
contraction velocity as one at a different amplitude $\epsilon^\prime$ and 
nondimensional time \hbox{${\tilde t}^\prime\,=\,
{\tilde t}\left(\epsilon/\epsilon^\prime\right)^2$}. The nondimensional time 
scales as the inverse square of the amplitude because cloud contraction is 
driven by a second-order perturbation. The dimensional time may be obtained 
from the nondimensional one through equation~(2). Does this scaling relation
mean that line profiles are degenerate in $\epsilon$ and $t$? Armed with 
RATRAN, we now demonstrate that this is fortunately not the case.
   
Figure~4 shows in detail the effect of trading amplitude and age. The solid
curve in the lefthand panel is the velocity profile in our canonical model
(\hbox{$\epsilon\,=\,0.1$}; \hbox{$t\,=\,1\,\,{\rm Myr}$}). The dashed curve is
the profile for a cloud with the same mass of \hbox{$3\,\Msun$}, but with
\hbox{$\epsilon\,=\,0.2$} and \hbox{$t\,=\,0.25\,\,{\rm Myr}$}. As predicted,
the velocities are quite similar. On the other hand, the internal
{\it densities} differ greatly. The righthand panel shows that the cloud with
larger $\epsilon$ has a lower central density, since it was initially more
inflated and evolved over a shorter time.

The net result is that the line profiles also do not match. As an example,
Figure~5 displays the three hyperfine lines of HCN. The degree of blue/red
asymmetry in all the profiles is markedly less than in the canonical model.
For the optically thinnest, $F\,(0-1)$ line, the profile is nearly Gaussian. 
The normalized velocity shift $\delta V$ is only
\hbox{$-0.16\,\,{\rm km}\,\,{\rm s}^{-1}$}, well below typical observed values.

What happens if we keep \hbox{$\epsilon\,=\,0.2$}, but increase $t$ back to
\hbox{$1\,\,{\rm Myr}$}? For our model to be useful observationally, the 
resulting line profiles should again differ noticeably from the canonical case.
Indeed they do, and in the opposite sense as the previous example. Both the CS
and HCN profiles now exhibit {\it too much} asymmetry. For example, the red
wing of the $F\,(0-1)$ line of HCN degenerates to a shoulder; a similar profile
is seen in CS. According to Figure~1 of \citet{me00}, this development reflects
relatively high infall speed. In fact, the maximum infall speed for this choice
of parameters is 0.06~km~s$^{-1}$, three times that in the canonical one. The 
central density is also higher by about the same factor. Consequently, the 
N$_2$H$^+$ line, while maintaining a Gaussian profile, has a $T_A$-value of 
0.85~K, significantly above the range reported by \citet{lmt99}.

\subsection{Abundance Gradients}

We have seen that the calculated CS profile is not even qualitatively correct 
unless we include the effect of depletion onto grain surfaces. Our strategy, 
following previous authors, was simply to introduce a spatial variation in the
molecular abundance. Nevertheless, our canonical model still assumes a uniform
distribution of N$_2$H$^+$ and HCN. This simplification also bears 
examination. 

The detailed abundance of any species is influenced not only by the efficiency
of grain absorption but also by the cloud dynamics itself. The latter point 
was emphasized in the pioneering study by \citet{ry01}, who followed a 
time-dependent chemical reaction network in models of collapsing dense cores.
Utilizing a radiative transfer code, these authors demonstrated, as we did in
\S 4.2, the sensitivity of emergent molecular line profiles to the cloud's 
internal velocity structure. Not only is there a direct, kinematic effect, but
various species may not have time to achieve chemical equilibrium, creating
abundance gradients that also influence line profiles. In a more recent work,
\citet{t08} found that the fractional abundance of N$_2$H$^+$ falls by an order
of magnitude from cloud center to edge within a cloud undergoing inside-out 
collapse (see their Fig.~3).

To account fully for the interplay of dynamics and chemistry, we would also 
need to follow simultaneously both a chemical reaction network and line 
transfer in our evolving cloud model. Needless to say, such an undertaking is
beyond the scope of this study. However, we may assess in a general way the 
effect of abundance gradients. Consider the optically thick lines of HCN, whose
profiles should be especially sensitive. In the spirit of \citet{t08}, we 
adopt a linear variation of the relative HCN abundance:
\begin{equation}
X\,=\,X_\circ\,\left[1\,+\,\beta\,\left(M_r/M\,-\,1/2\right)\right] \,\,.
\end{equation}
Here, $M_r$ is the mass contained within any radius $r$ and $X_\circ$ is the
nominal abundance we used previously. Within our canonical model, we vary the
nondimensional slope $\beta$ and calculate the emergent HCN profiles.

The slope $\beta$ cannot exceed 2 in absolute value, since $X$ would then
become negative at either the cloud center or edge. For smaller, but 
substantial, values of $\vert\beta\vert$, the most significant effect is on
the depth of the self-absorption dip. Figure~6 shows how the optically
thickest, $F\,(2 - 1)$, line profile changes as $\beta$ ranges from -0.5 
through 0 (uniform abundance), to +0.5. According to the standard gauge of the
relative velocity shift, the effect is minor. For \hbox{$\beta\,=\,-0.5$},
$\delta V_{\rm HCN}$ is 6~\% greater in magnitude than the canonical value, 
while the shift is 6~\% less for \hbox{$\beta\,=\,+0.5$}. 

Nevertheless, the degree of self-absorption diminishes markedly for positive 
$\beta$, i.e., when HCN is concentrated in the more slowly moving, outer 
portion of the cloud. To restore the dip to its original magnitude, the
contraction age for the cloud would need to increase slightly from the 
canonical 1.0~Myr to 1.2~Myr. We conclude that, while coupled chemical and 
dynamical models are preferable in principle, such extensive effort is 
unnecessary for beginning a quantitative assessment of starless core 
properties.

\section{Discussion}

This suite of radiative transfer calculations performed on our theoretical 
cloud model has yielded two important findings. First, we have been able to 
match the generic, observed properties of the selected line profiles using a 
reasonable cloud mass, turbulent velocity dispersion, and evolutionary time. 
This success may indicate that the model has captured the essential physics 
of the contraction process itself. Note also that the accelerating character of
the contraction in our theory explains naturally the key observation that 
starless cores of lower density exhibit less asymmetry in their line profiles 
\citep{ge00,wi06}. Second, the sensitivity of the emergent line profiles to 
the input parameters means that the model offers a practical route to gauge 
starless core properties that have previously been difficult to ascertain. For 
example, given the empirical mass and kinetic temperature of an object, the 
matching of line profiles would yield a first estimate of its contraction age. 
Obtaining ages for numerous starless cores in different environments would aid 
greatly in understanding the onset of star formation. 

Of course, the observations offer more detail than the model can explain, at
least in its current, highly idealized form. We cannot yet account for the
rather surprising spatial extent of asymmetric CS emission \citep{me00,lmt01}.
Nor can we explain the relative intensities of the HCN hyperfine lines
\citep{s04}. Explaining both features of the data may require alterations in
the spatial abundance gradients, in the manner employed in Sections 4.2 and 
4.4. Additionally, an improved model would consider more carefully how the
interior, relatively quiescent gas joins onto surrounding, more turbulent, 
material. In fact, it has long been realized that the pattern of HCN
hyperfine anomalies requires the presence of an envelope of relatively high 
excitation temperature \citep{g93}. \citet{go09} have recently simulated the
birth of dense cores out of spherical, converging flows. Such calculations,
when extended to three dimensions, may offer additional insight into the 
nature of the core-envelope transition.  

Finally, we must face the obvious fact that real starless cores are not
spherical, as the model currently assumes. The shapes of these objects may be
partially determined by the anisotropic pressure of their turbulent 
environments, as found in recent simulations \citep[e.g.][]{o08}. Another 
important factor setting the gross core properties is the force associated with
the internal magnetic field \citep{l04}. It would be instructive to study, 
using the method of Paper~I, the onset of collapse in marginally stable, 
magnetostatic equilibria, such as those found by \citet{t88}. Such an 
investigation would add a significant degree of realism to our picture of 
protostellar collapse. 

\acknowledgments

We are grateful to Floris van der Tak, one of the authors of RATRAN, for
guiding us in the proper use of the code. Jorge Pi\~neda, who is independently
analyzing the line emission from cloud models, also provided much-valued
assistance, as did Melissa Enoch. Throughout the project, S.~S. was partially
supported by NSF grant AST-0908573. 

\clearpage

\appendix

\clearpage

\begin{figure}
\plotone{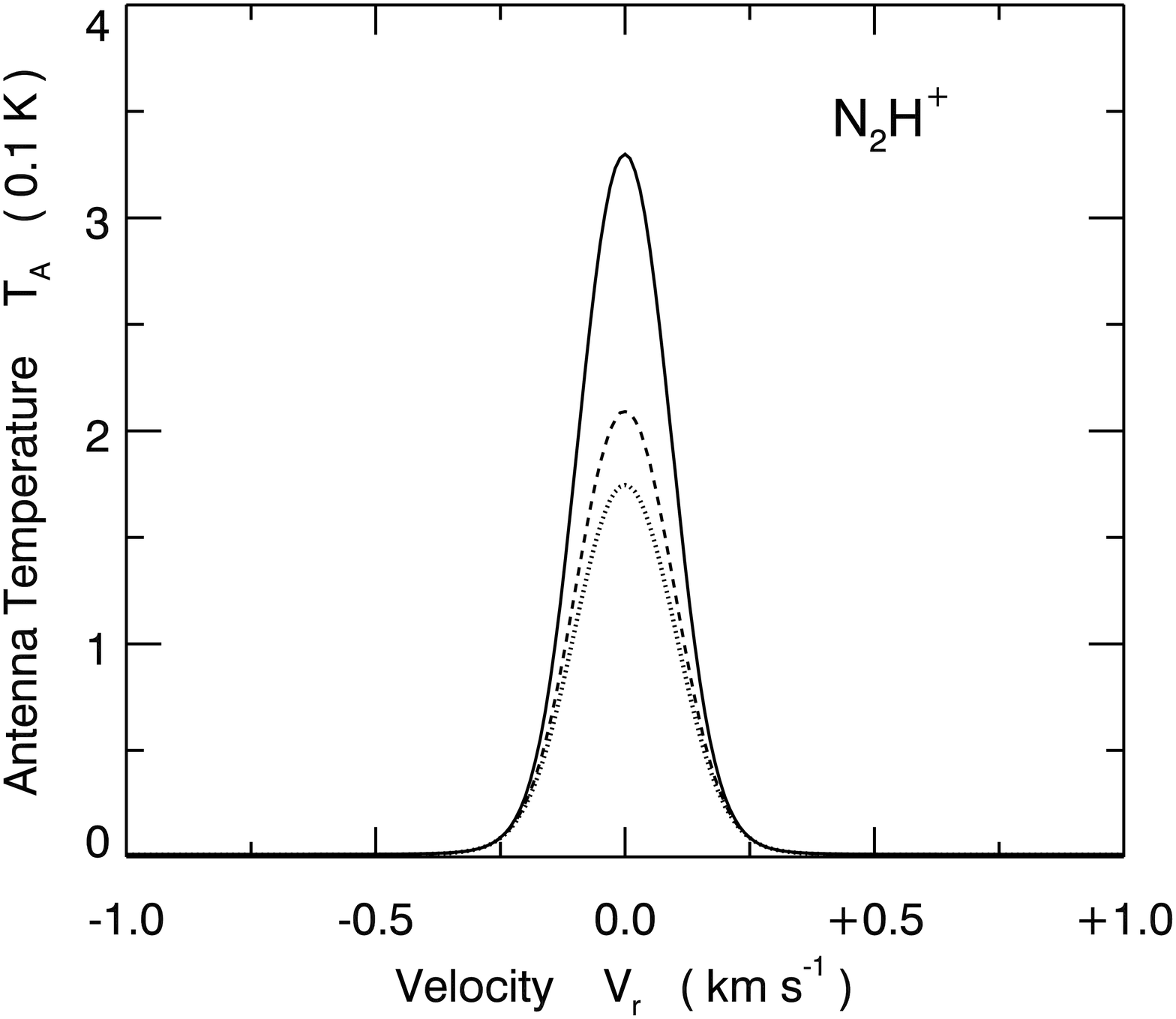}
\caption{Profile evolution of the 
\hbox{${F_1 ,F}\,=\,{0,1} \rightarrow {1,2}$} hyperfine line within the
\hbox{$J\,=\,1\,\rightarrow\,0$} rotational transition of N$_2$H$^+$. The
profiles, viewed through the cloud center, are for a cloud of $3\,\,\Msun$
with \hbox{$\kappa\,=\,5.5\times 10^{-12}\,\,{\rm g}\,\,{\rm cm}^{-1}\,\,
{\rm s}^{-2}$}. From bottom to top, the times are 0, 0.5, and 1.0~Myr.}
\end{figure}

\begin{figure}
\plotone{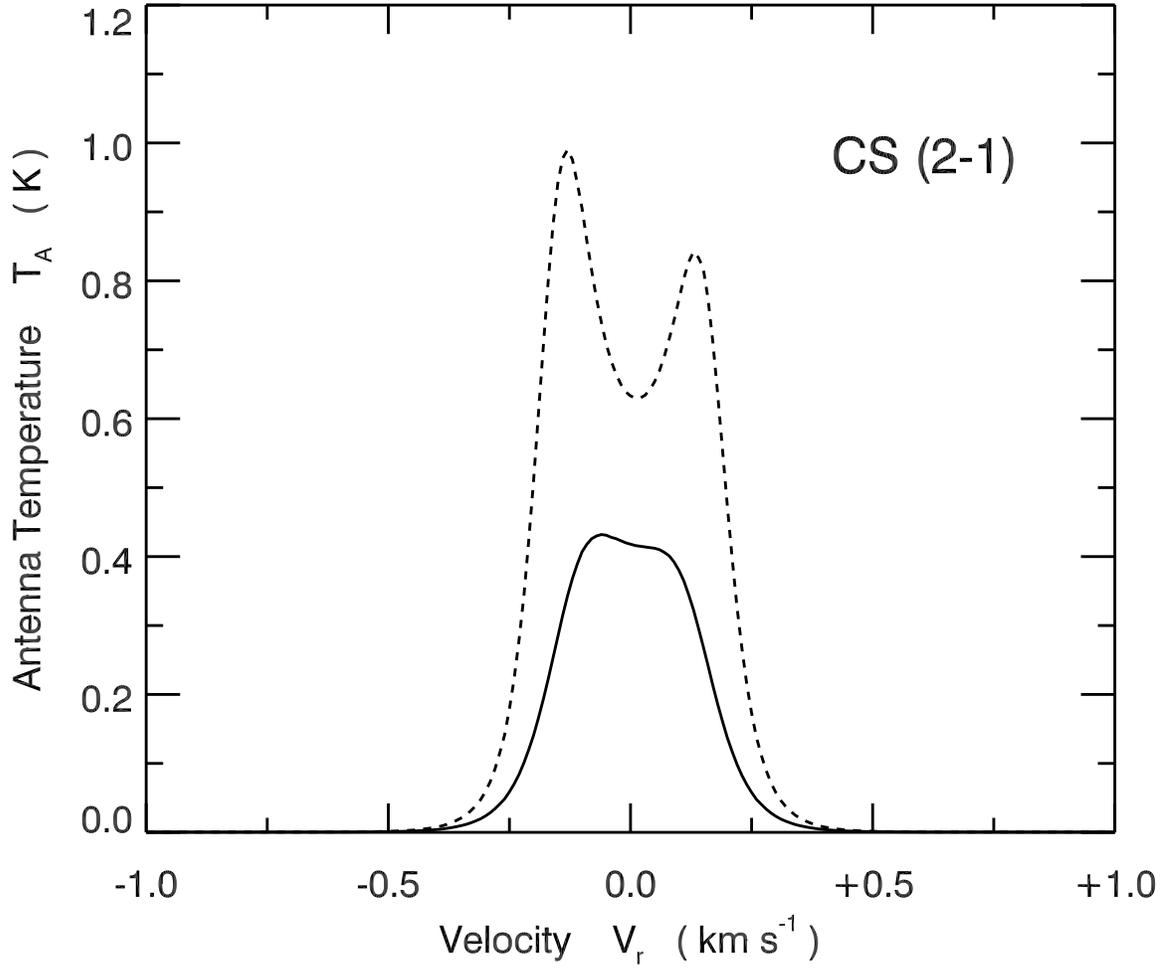}
\caption{Profile of the \hbox{$J\,=\,1\rightarrow 0$} rotational transition
of CS, viewed through the cloud center for the canonical cloud model. The
dashed curve is the profile using a uniform CS relative abundance of
\hbox{$X\,({\rm CS})\,=\,4\times 10^{-9}$}. For the solid profile, the 
molecule was centrally depleted, following the prescription of \citet{t02}.}
\end{figure}

\begin{figure}
\plotone{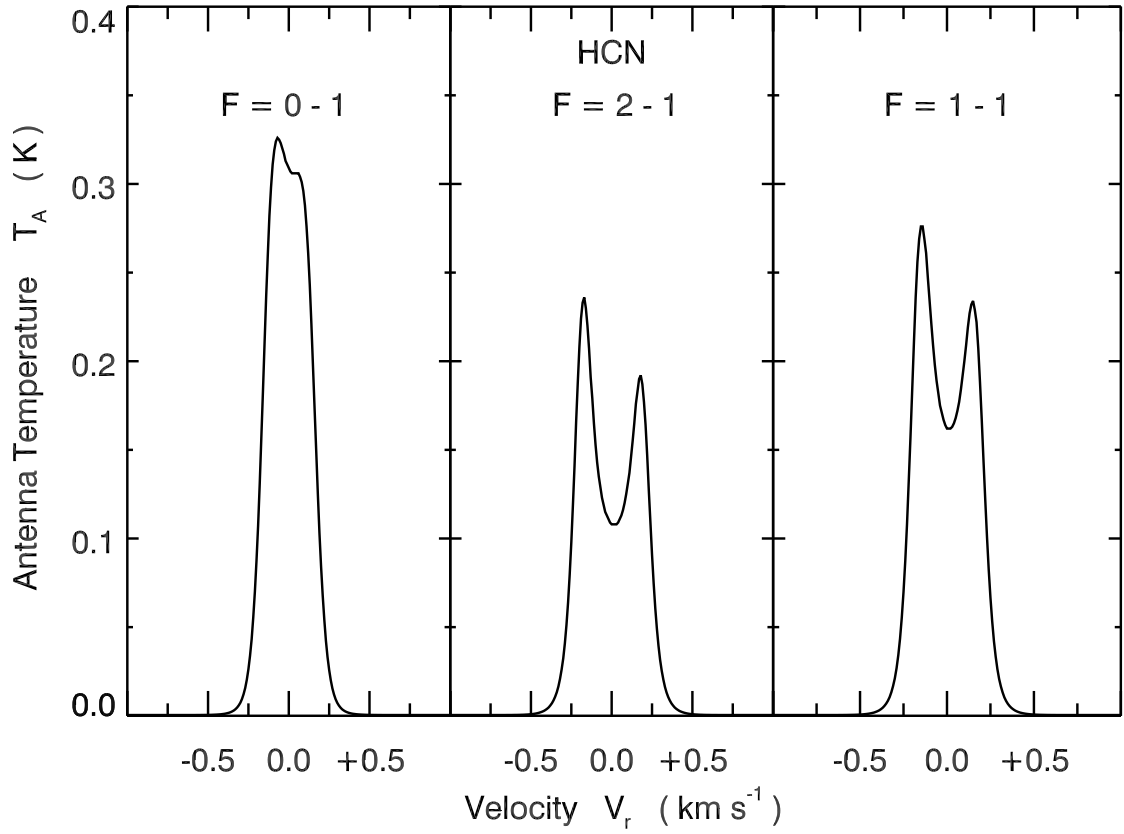}
\caption{Profiles of HCN for the three hyperfine lines within the
\hbox{$J\,=\,1\rightarrow 0$} rotational transition. All profiles are
calculated using the canonical model, and are viewed through the cloud center.}
\end{figure}

\begin{figure}
\plotone{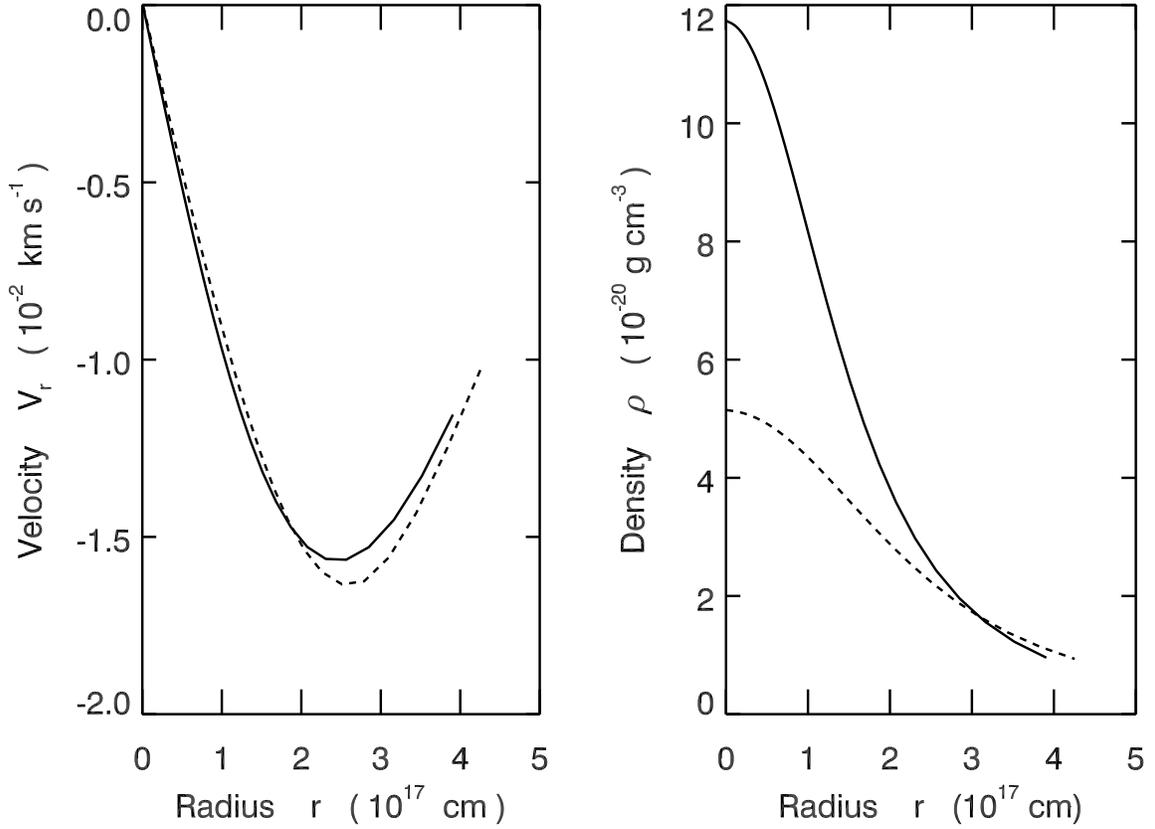}
\caption{Theoretical velocity and density profiles. The left panel shows the
contraction velocity as a function of radius. The solid curve is the canonical
model (\hbox{$\epsilon\,=\,0.1$}, \hbox{$t\,=\,1\,\,{\rm Myr}$}), while the
dashed curve has \hbox{$\epsilon\,=\,0.2$} and 
\hbox{$t\,=\,0.25\,\,{\rm Myr}$}. The right panel displays the density 
profiles for the same two models.}
\end{figure}

\begin{figure}
\plotone{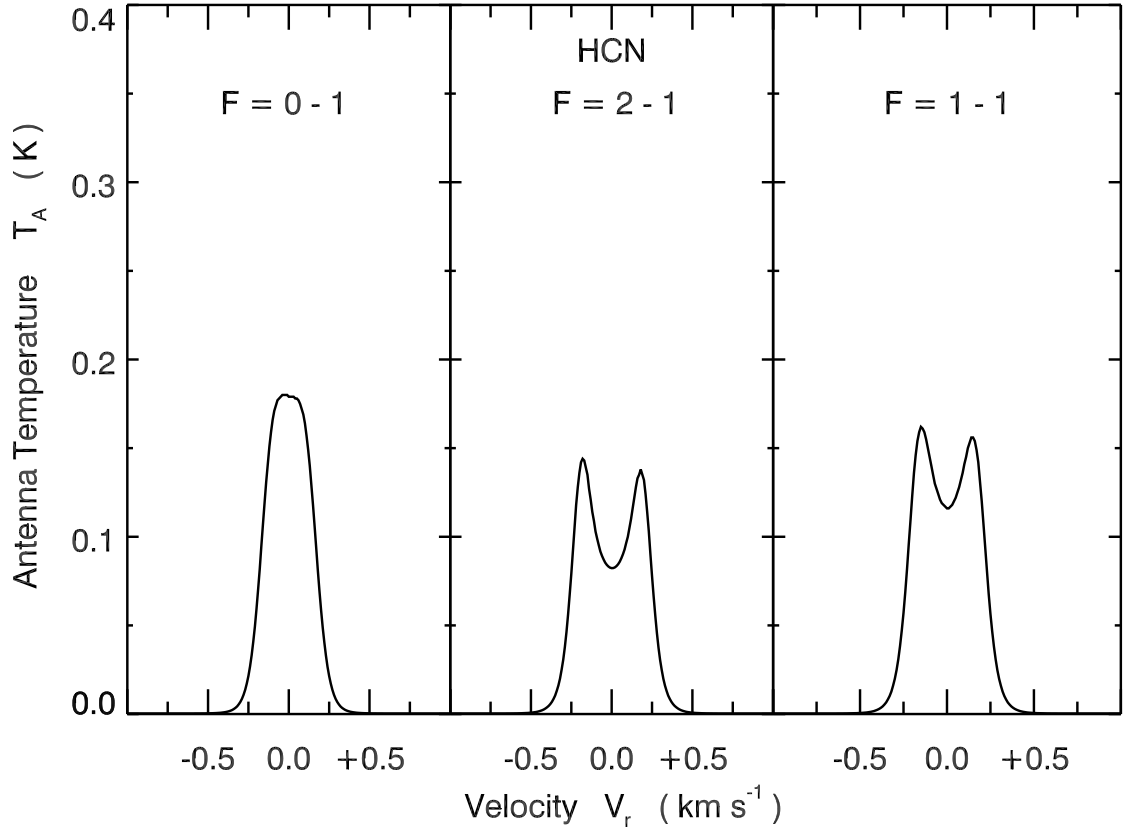}
\caption{The effect of changing $\epsilon$ and $t$ on the three hyperfine 
lines of HCN. All profiles are calculated using \hbox{$\epsilon\,=\,0.2$} and 
\hbox{$t\,=\,0.25\,\,{\rm Myr}$}, and are viewed through the cloud center. 
Note how the asymmetry is decreased with respect to the profiles of Figure~3.} 
\end{figure}

\begin{figure}
\plotone{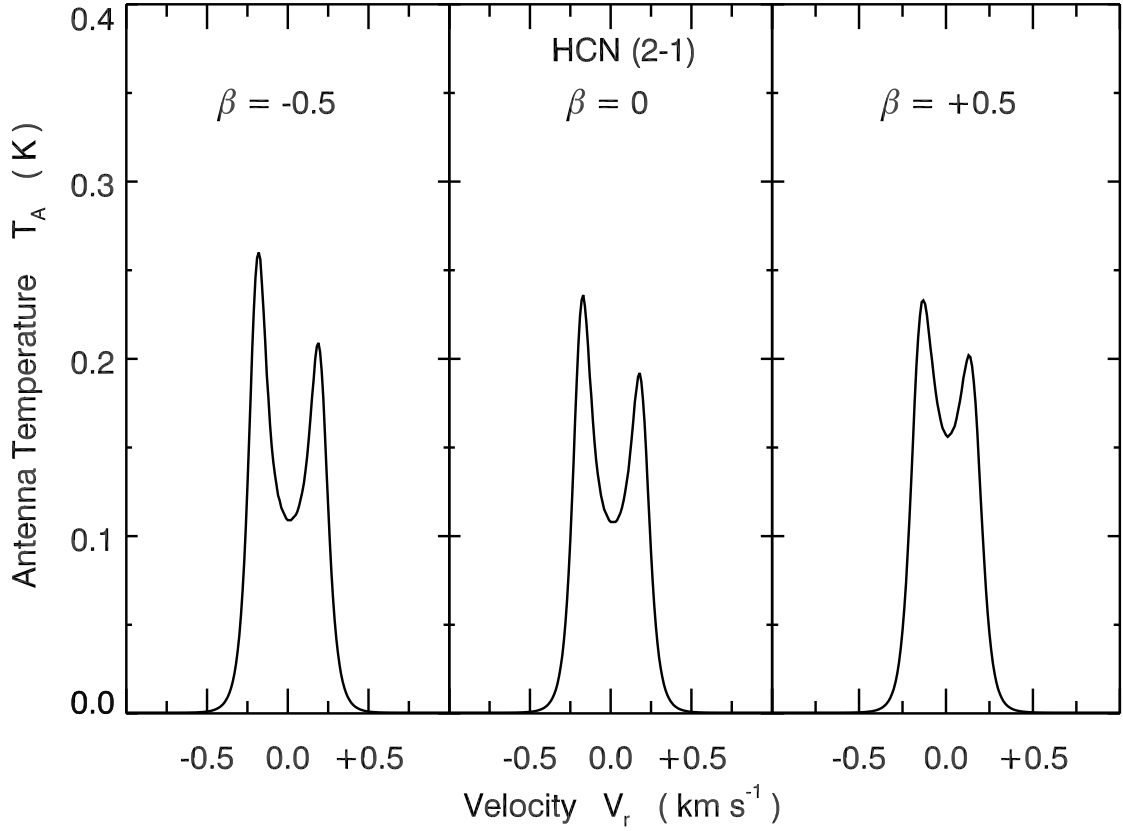}
\caption{The effect of abundance gradients on the $F\,(2-1)$ HCN hyperfine 
line. The cloud model is the canonical one, with the addition of a linear HCN 
abundance gradient. The slope $\beta$, as defined in equation~(6) is given in 
each panel. Notice how the self-absorption dip decreases for positive 
$\beta$.}   
\end{figure}

\end{document}